\documentclass{iau} 
\usepackage{graphicx}
\usepackage{natbib}
\usepackage{hyperref}

\title[PION: Simulations of Wind-Blown Nebulae] 
{PION: Simulations of Wind-Blown Nebulae}

\author[Jonathan Mackey \textit{et al.}]   
{Jonathan Mackey$^{1,2}$,
 Samuel Green$^{1,2}$,
 Maria Moutzouri$^{1,2}$,
 Thomas J.~Haworth$^{3}$,
 Robert D.~Kavanagh$^{4}$,
 Maggie Celeste$^{1,5}$,
 Robert Brose$^{1,2}$,
 Davit Zargaryan$^{1,2}$
 \and Ciar\'an O'Rourke$^{6}$}

\affiliation{$^1$Dublin Institute for Advanced Studies, 31 Fitzwilliam Place, Dublin 2, Ireland
\\[\affilskip]
$^{2}$DIAS Dunsink Observatory, Dunsink Lane, D15 XR2R, Ireland
\\[\affilskip]
$^3$Astronomy Unit, School of Physics and Astronomy, Queen Mary University of London, London E1 4NS, UK
\\[\affilskip]
$^4$Leiden Observatory, Leiden University, PO Box 9513, 2300 RA, Leiden, The Netherlands
\\[\affilskip]
$^5$School of Physics, Trinity College Dublin, The University of Dublin, Dublin 2, Ireland
\\[\affilskip]
$^6$Irish Centre for High-End Computing (ICHEC), NUI Galway, Galway City, Ireland
}

\pubyear{2022}
\volume{362}  
\jname{The predictive power of computational astrophysics as a discovery tool}
\editors{D.~Bisikalo, T.~Hanawa, C.~Boily \& J.~Stone, eds.}
\begin{document}

\maketitle

\begin{abstract}
We present an overview of PION, an open-source software project for solving radiation-magnetohydrodynamics equations on a nested grid, aimed at modelling asymmetric nebulae around massive stars.
A new implementation of hybrid OpenMP/MPI parallel algorithms is briefly introduced, and improved scaling is demonstrated compared with the current release version.
Three-dimensional simulations of an expanding nebula around a Wolf-Rayet star are then presented and analysed, similar to previous 2D simulations in the literature.
The evolution of the emission measure of the gas and the X-ray surface brightness are calculated as a function of time, and some qualitative comparison with observations is made.
\end{abstract}

\firstsection 
\section{Introduction}

The late stages of massive-star evolution, after the main sequence, remain very uncertain \citep{Lan12, Smi14}, despite significant advances in recent years.
In particular, uncertainties in the mechanisms of mass loss of Red Supergiants (RSG) and Luminous Blue Variables (LBV) make it difficult to predict whether a given star can lose its envelope through mass loss during its lifetime and become a Wolf-Rayet (WR) star.
Mass transfer and mass loss induced by interaction with a binary companion is common in the evolution of massive stars \citep{SanDeMdeK12}, and so the relative importance of mass loss through winds and eruptions, versus mass-loss through binary interaction, is an active field of research.
This lack of knowledge translates to a large systematic uncertainty when making predictions for evolutionary tracks of massive stars and for connecting observed properties of supernova progenitors with the zero-age main-sequence progenitor properties (e.g.\ mass).

Many post-main-sequence massive stars are surrounded by nebulae that can be composed of mass lost during previous evolution and/or interstellar gas swept up by the expanding stellar wind \citep{GarLanMac96}.
The dynamical timescale of these nebulae is $\sim10^4-10^5$ years, not much shorter than the $\sim10^5-10^6$ year timescales of post-main-sequence nuclear burning in massive stars.
Constraints on the previous evolution of stars can therefore be obtained by comparing predictions of stellar evolution calculations with observed nebulae \citep{MacMohNei12, MeyPetPoh20}.
This is not a trivial undertaking because the shocks driven by expanding wind bubbles are often radiative and subject to dynamical instabilities \citep{GarMac95b}, and effects of thermal conduction \citep{ComKap98, MeyMacLan14} and magnetic fields \citep{VanMelMar15, MeyMigKui17} can be significant.
Multi-dimensional simulations are often required to capture the effects of these processes on the evolution of a circumstellar nebula.

\begin{figure}[ht]
\begin{center}
 \includegraphics[width=3.4in]{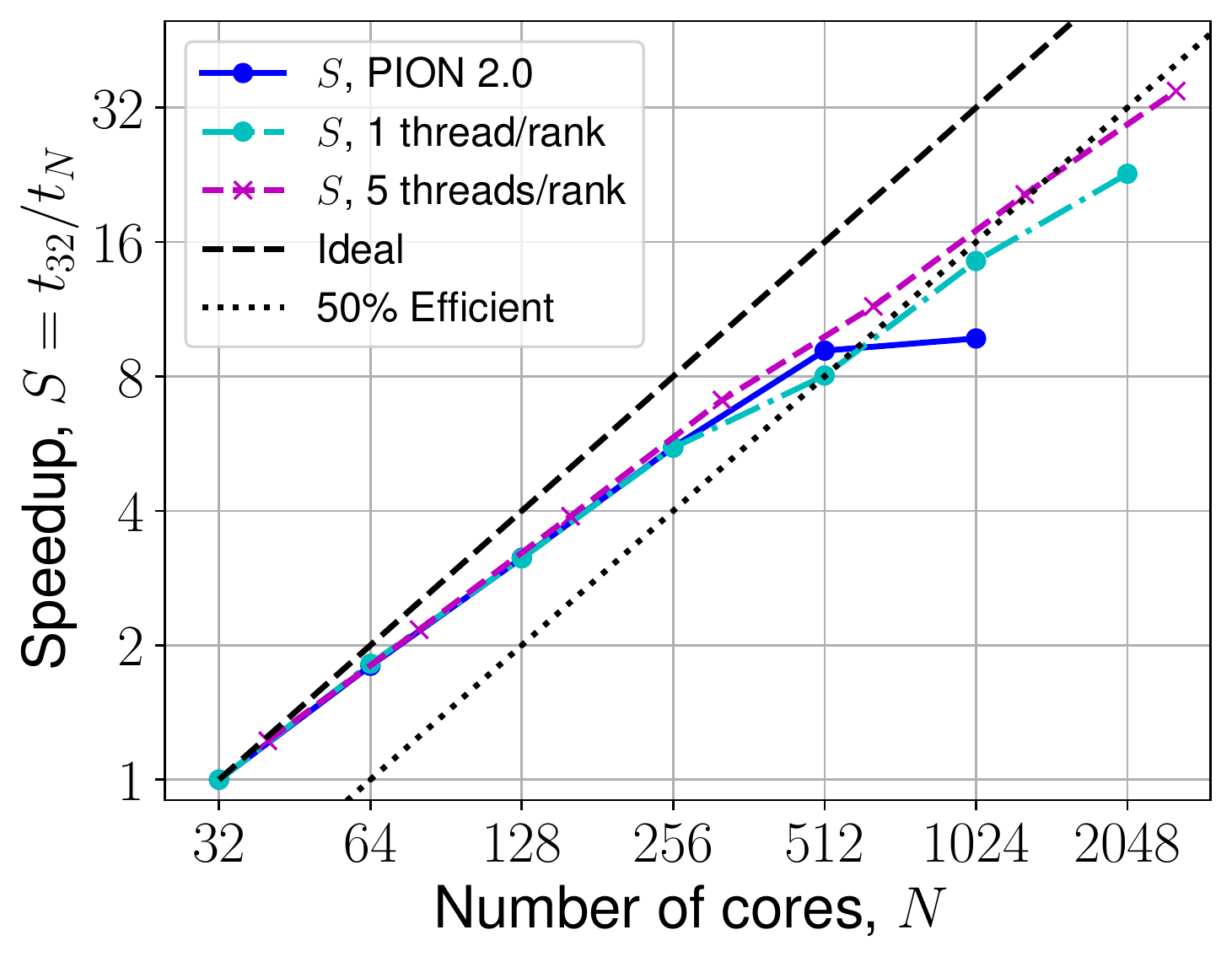} 
 \caption{Strong scaling of PION for a 3D MHD simulation of a bow shock around a runaway massive star.  The simulation has $256^3$ grid cells per level and 3 levels of refinement.  Results from PION v2.0 from \cite{MacGreMou21} (blue solid line) are compared with an upgraded version of PION run with 1 (cyan dot-dashed line) and 5 (magenta dashed line) OpenMP threads per MPI process.  The speedup, $S$, is defined as the run duration using $N$ cores, $t_N$, divided by the run duration using 32 cores, $t_{32}$.}
   \label{fig:scaling}
\end{center}
\end{figure}

\section{Methods}
In \citet{MacGreMou21} we presented the first public release of PION, a radiation-magnetohydrodynamics (R-MHD) code for simulating the circumstellar medium (CSM) around massive stars and, more generally, feedback of massive stars on the interstellar medium (ISM).
PION uses the finite-volume method on a rectilinear grid of cubic cells to solve the equations of hydrodynamics or ideal-MHD following the time-integration scheme of \citet{FalKomJoa98}.
A short-characteristics raytracing module is used to calculate photon fluxes at each cell, that can be used to obtain non-equilibrium ionization, heating and cooling rates \citep{Mac12}.
Nested grids (i.e.\ static mesh-refinement) focus resolution on certain points in the domain, usually on the star for the case of wind-blown nebulae \citep[cf.][]{YorKai95,FreHenYor03}.
PION also has stellar-wind boundary conditions for non-rotating and rotating, evolving stars, following methods developed for Heliosphere modelling \citep{PogZanOgi04}.
The source code and tutorials for PION v2.0 are available from a website\footnote{\href{https://www.pion.ie/}{https://www.pion.ie}} and a VCS repository\footnote{\href{https://git.dias.ie/massive-stars-software}{https://git.dias.ie/massive-stars-software}} with a BSD 3-clause license.

PION is written in C++ and is parallelised using domain decomposition into sub-domains with the Message Passing Interface (MPI).
Since v2.0 was released we have worked on improving parallel scaling to larger number of cores, by upgrading the MPI communication algorithms and implementing hybrid OpenMP/MPI parallelisation.
The idea behind this is that the ratio of (duplicated) boundary data to domain data increases with the number of sub-domains (PION allocates one sub-domain per MPI process on each level of the nested grid), and this is the fundamental limit to the parallel scaling.
If we reduce the number of MPI processes for a given core count, by using OpenMP threads, then the parallel scaling should improve.

It was relatively straightforward to rewrite the loops over cells within each sub-domain so that they are loops over 1D columns of cells.
Each OpenMP thread is then given a 1D column of cells to calculate on for the various tasks (calculation of timestep, hydrodynamic fluxes in each direction, radiative heating and cooling, ionization and recombination, and cell updates).
The most difficult work involved making various function calls threadsafe for OpenMP execution, which required some re-design of the data structures.
Code optimization work is still ongoing, but there is already signficiant improvement.

Fig.~\ref{fig:scaling} shows preliminary results of the strong scaling using upgraded MPI communication and either 1 or 5 OpenMP threads per MPI process.
The speedup achieved with respect to a simulation using 32 cores is compared with the ideal scaling.
The scaling that was achieved using the same simulation in \citet{MacGreMou21} is shown for comparison, and it is clear that we have made significant progress.
Using 5 threads per rank we are able to achieve good scaling up to at least 2560 cores, $5\times$ as many cores as previously.
The speedup has still not saturated at this number of cores, for a test calculation with $256^3$ cells per level and 3 refinement levels.
This gives us confidence that we could achieve good scaling to 10\,000 cores or more on larger grids of, e.g., $512^3$ or $1024^3$ cells per refinement level.
These scaling improvements are enabled by our participation in the EuroCC-Ireland Academic Flagship Programme, and will be included in the next release of PION.

Synthetic emission maps for some observational tracers (emission measure, broadband X-ray emission, H$\alpha$) can be generated with a projection code distributed with PION, which stores the projected maps in binary VTK format.
A python library is provided that can read both PION snapshots and the VTK projected-data files and generate simple plots.
For calculating infrared dust emission, PION snapshots are exported to the TORUS Monte-Carlo radiative transfer code \citep{HarHawAcr19}.
This then determines the spatially varying dust temperature via a radiative equilibrium calcuation, and produces emission maps of the thermally emitting dust grains at specified frequencies \citep{GreMacHaw19}.

\section{Wolf-Rayet Nebulae}
Nebulae around WR stars are often spectacular objects, as the intense ionizing radiation and dense, fast winds of the stellar core sweep through the remnants of the extended stellar envelope, lost either through binary stripping or winds/eruptions.
Good examples are the nebula M1-67 around WR\,124 \citep{GroMofJon98} and NGC\,3199 around WR\,18 \citep{ToaMarGue17}.
The first multidimensional calculations of this process by \citet{GarLanMac96} followed the CSM around a 35\,M$_\odot$ star as it evolved from a O star $\rightarrow$ RSG $\rightarrow$ WR, using 1D stellar-evolution calculations as a time-dependent inner boundary condition to 2D hydrodynamical simulations in spherical geometry.
Later the same evolutionary calculation was used in radiation-hydrodynamical (R-HD) simulations using a 2D nested grid in cylindrical coordinates \citep{FreHenYor06}.

In \citet{MacGreMou21} we again used the same evolutionary calculation on a 3D Cartesian nested grid to study the performance of PION on a calculation where there are 2D reference results in the literature.
The grid has $256^3$ cells on each level and 4 levels, the coarsest level being a cube with domain $\{x,y,z\} \in [-30,30]\times10^{18}$\,cm and each finer level $2\times$ smaller in each dimension and centred on the star located at the origin.
The wind boundary region has a radius of 20 cells, $\approx 5.86\times10^{17}$\,cm.
The time evolution of the stellar wind and radiation, as well as snapshots showing slices through the 3D simulation domain, are presented in \citet{MacGreMou21}.

\begin{figure}
\begin{center}
 \includegraphics[width=0.48\textwidth]{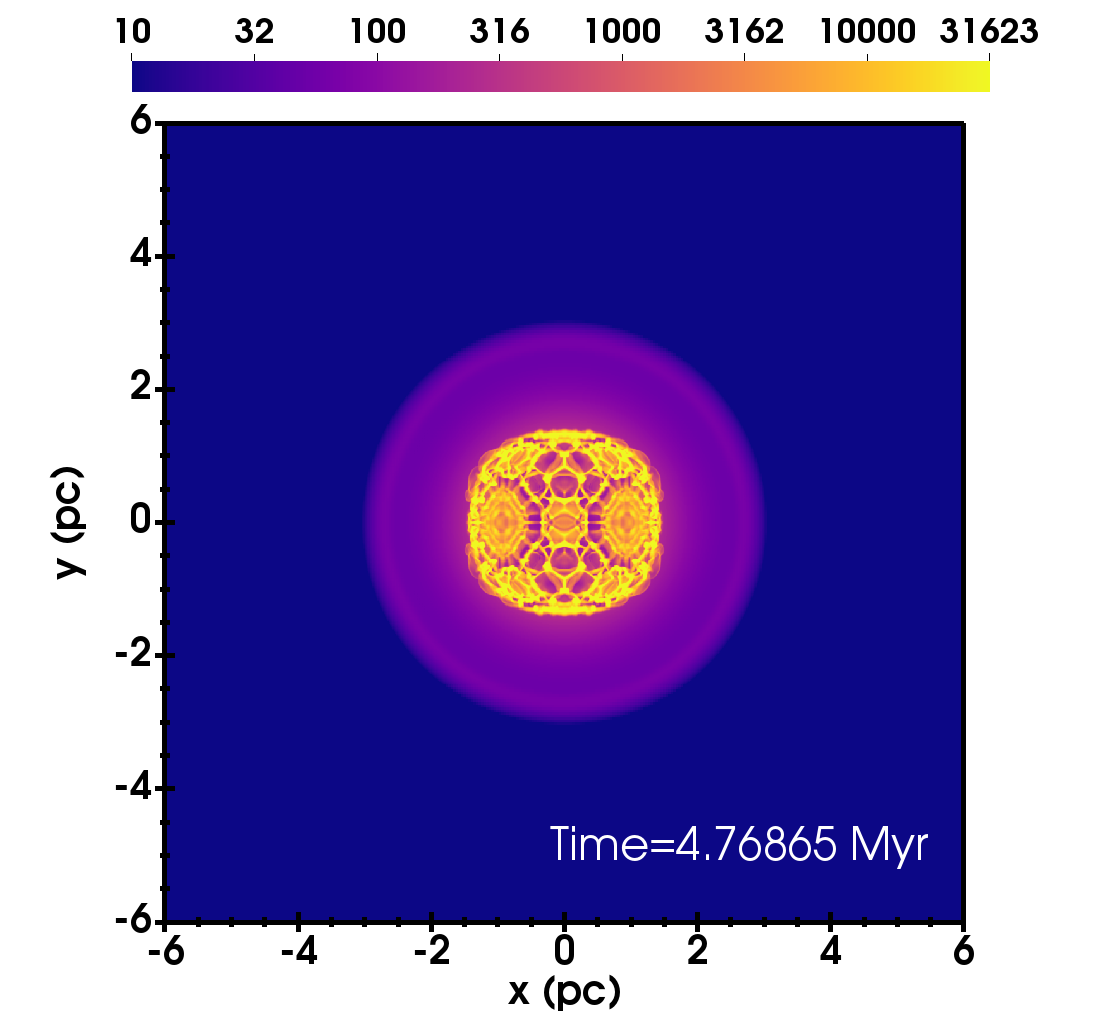} 
 \includegraphics[width=0.48\textwidth]{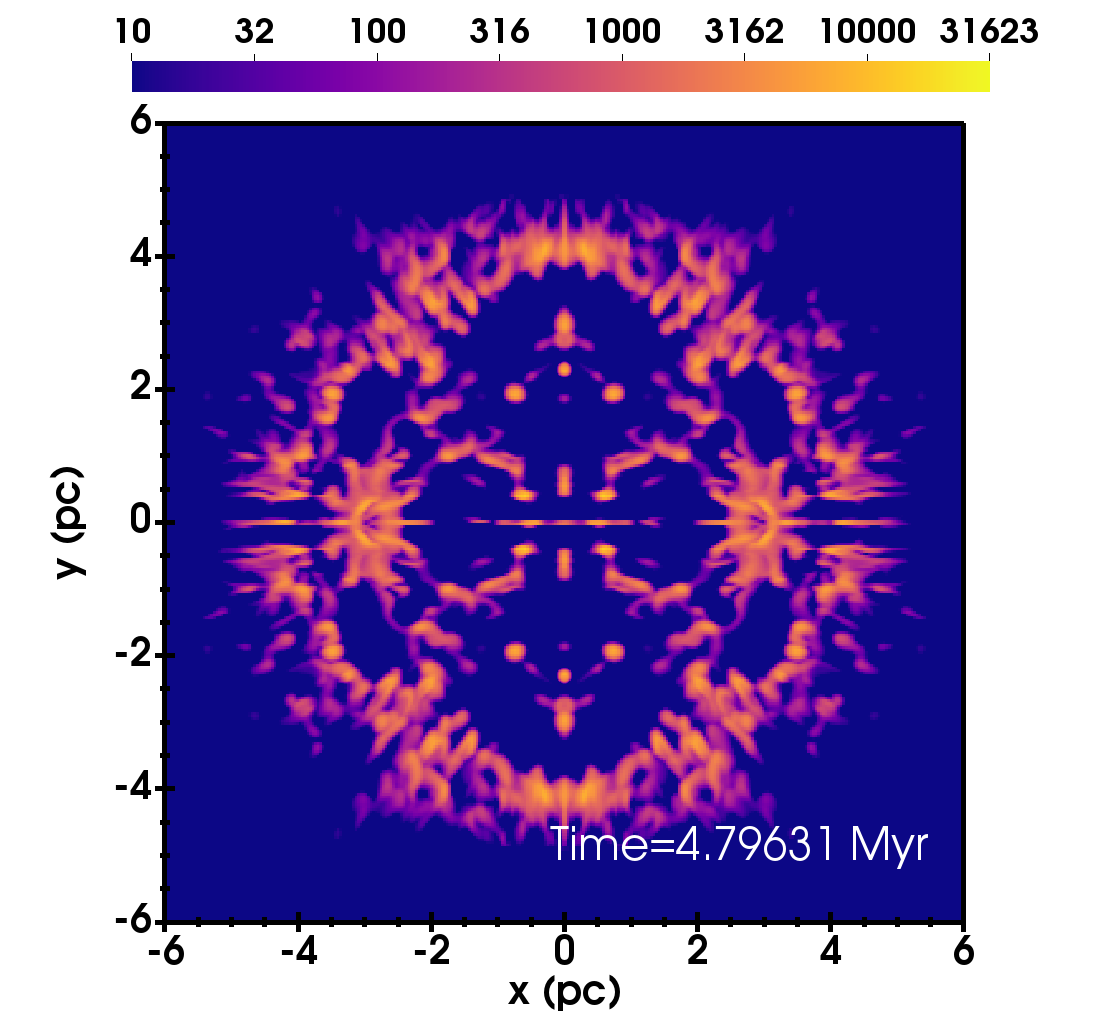} 
 \caption{Emission measure from the expanding nebula 14\,000 years (left) and 42\,000 years (right) after the RSG$\rightarrow$WR transition, plotted on a logarithmic scale with units cm$^{-6}$\,pc. Time is shown since the beginning of the stellar evolution calculation; the RSG phase ended at $t\approx4.755$\,Myr.}
   \label{fig:EM}
\end{center}
\end{figure}

Here we present synthetic observations of the simulation, calculating the projected Emission Measure (EM) and X-ray surface brightness using a raytracing method, neglecting internal absorption.
The EM is plotted in Fig.~\ref{fig:EM} as the fast wind from the WR star sweeps up the wind bubble of the previous RSG phase.
The slow and dense wind from the RSG expanded to distance from the star (located at the origin) of $r\approx3$\,pc, and the WR wind-bubble expands through this dense gas at about $120$\,km\,s$^{-1}$.
The wind termination shock is adiabatic, but the forward shock is strongly radiative leading to dynamical instability in the swept-up shell.
14\,000 years after the RSG$\rightarrow$WR transition (at $t\approx4.755$\,Myr), the shell has expanded to 1.7\,pc and remains relatively regular and symmetric.
Once the swept-up shell breaks out of the RSG wind region it breaks up into clumps and filaments, and 42\,000 years after the transition it is very fragmented and qualitatively similar in appearance to Galactic WR nebulae like M1-67.

\begin{figure}
\begin{center}
 \includegraphics[width=0.48\textwidth]{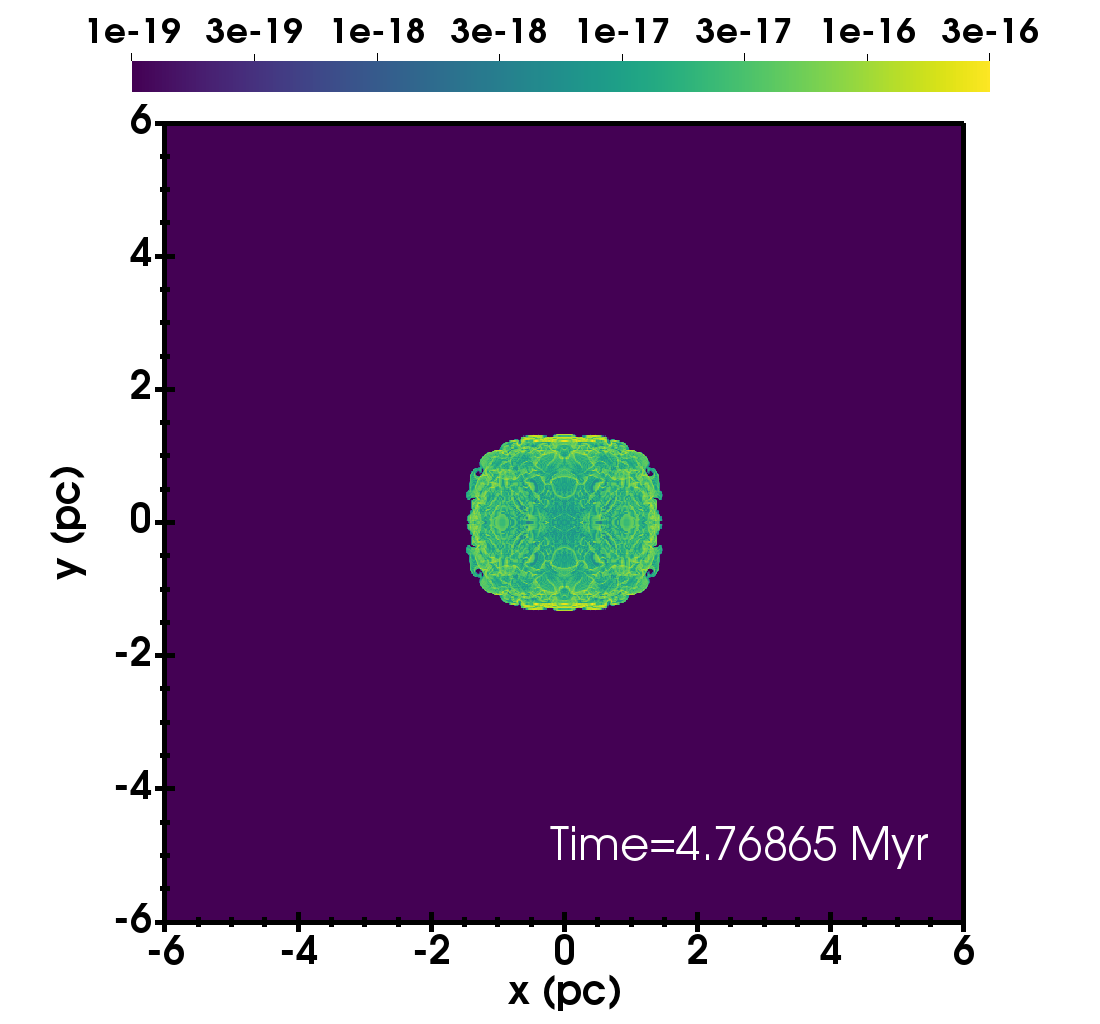} 
 \includegraphics[width=0.48\textwidth]{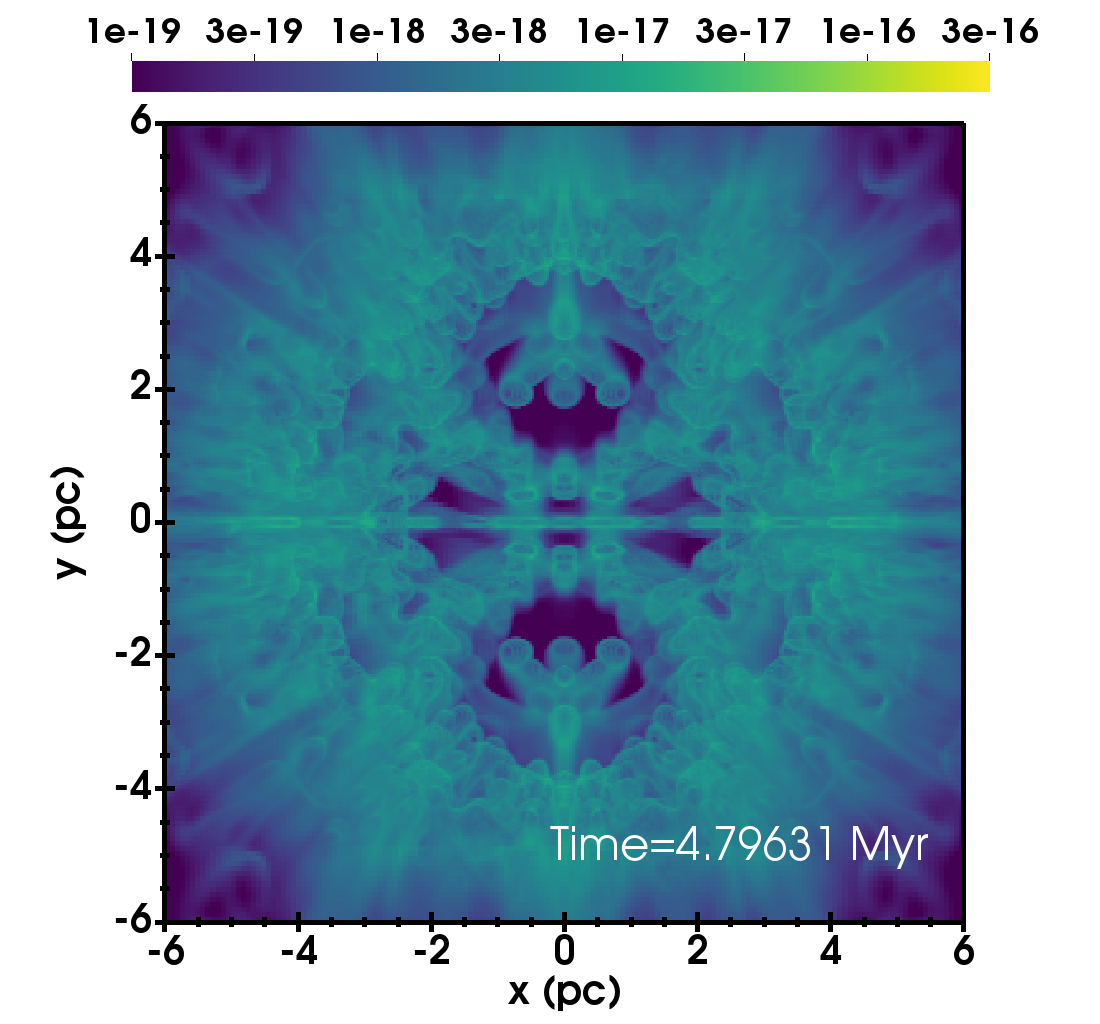} 
 \caption{X-ray emission (0.3-10\,keV) from the WR nebula 14\,000 years (left) and 42\,000 years (right) after the RSG$\rightarrow$WR transition, plotted on a logarithmic scale with units erg\,cm$^{-2}$\,s$^{-1}$\,arcsec$^{-2}$.}
   \label{fig:Xray}
\end{center}
\end{figure}

X-ray emission from 0.3-10\,keV is shown in Fig.~\ref{fig:Xray}, calculated using the method in \citet{GreMacHaw19}, plotted at the same times as for the EM.
When the expanding WR wind is still contained within the RSG wind region the X-ray emission is bright and fills the WR wind bubble, slightly limb-brightened.
At later times it is about $10\times$ fainter and associated with the boundary layers where hot coronal gas is dynamically mixing with photoionized (and much denser) nebular gas.
For both plots (EM and X-ray) the clumps and filaments have a very regular and symmetric appearance.
This is because the instabilities are seeded by the computational grid, and so the the resulting structures that develop reflect this cubic Cartesian grid.

In reality RSG winds are clumped, asymmetric and variable \citep{OGoVleRic15, HumDavRic21}, reflecting their turbulent outer convective layers, and so we should not expect detailed agreement between our simple model and any observed nebula.
Also, rotation was not included in this evolutionary calculation and this would introduce an asymmetry in the WR and RSG winds.
In future work we will investigate WR nebulae with higher resolution, 3D R-MHD simulations that are based on new stellar evolution sequences for single-star progenitors of classical WR stars.

\section{Conclusions}
PION is a 1D-3D R-MHD code for simulating expanding nebulae around stars -- the current public release is v2.0, available from \href{https://www.pion.ie/}{https://www.pion.ie}.
It has demonstrated parallel scaling to 1024 MPI processes on production simulations, as presented in \citet{MacGreMou21}, although the scaling for problems with radiative transfer is not so good.
Newly implemented hybrid OpenMP/MPI algorithms show much better scaling to at least 2560 cores on the same test problems, and these improvements will be included in the next public release of PION.
A 3D R-HD simulation of an expanding WR nebula was presented, including the time-evolution of projected Emission Measure maps and X-ray surface-brightness maps.
These show many features that are qualitatively similar to Galactic WR nebulae such as M1-67 and NGC\,3199, and motivate more detailed 3D simulations with the goal of determining which (if any) of the Galactic WR nebulae are produced by mass loss through stellar winds as predicted by single-star evolution.

\section*{Acknowledgements}
JM is grateful to N.~Langer and L.~Grassitelli for stimulating discussions and for providing evolutionary calculations.
JM acknowledges funding from a Royal Society-Science Foundation Ireland University Research Fellowship (20/RS-URF-R/3712).
DZ, RB acknowledge funding from an Irish Research Council (IRC) Starting Laureate Award (IRCLA\textbackslash 2017\textbackslash 83).
MM acknowledges funding from a Royal Society Research Fellows Enhancement Award (RGF\textbackslash EA\textbackslash 180214).
TJH is funded by a Royal Society Dorothy Hodgkin Fellowship. 
The authors wish to acknowledge the DJEI/DES/SFI/HEA Irish Centre for High-End Computing (ICHEC) for the provision of computational facilities and support (project eurocc-af-2).
This work is supported by the EuroCC project funded by the European High-Performance Computing Joint Undertaking (JU) under grant agreement No 951732 and the Irish Department of Further and Higher Education, Research, Innovation and Science.

\bibliographystyle{apj}
\bibliography{./refs}

\begin{discussion}

\discuss{Portegies Zwart}{
  The symmetries are beautiful but not very realistic.  One way to get rid of them is by introducing some random noise, but would they disappear if you increase the resolution and you get automatically randomness?
}

\discuss{Mackey}{
  It shouldn't -- it depends how symmetric the Riemann solver is.  If it is perfectly symmetric then all of the octants should be identical.  At the moment I am looking into adding a stochastic model for clumpiness to the RSG wind, because we know that the winds of cool stars are very clumped. We expect these non-linear clumps in the wind will then seed all of the structure that we see in the WR nebulae.
}

\discuss{Mohamed}{
  Could you say something about the computational cost of the TORUS post-processing of PION simulations.
}

\discuss{Mackey}{
  At the beginning the TORUS calculations were taking longer than the hydro simulations, but Tom and Sam did a lot of work on this.
  In the end with MPI and multi-threading, the TORUS calculations took about 10\,000 core hours for a 3D simulation that required about 50\,000 core hours.
}
\end{discussion}

\end{document}